\documentclass[conference]{IEEEtran}
\IEEEoverridecommandlockouts
\usepackage{cite}
\usepackage{amsmath,amssymb,amsfonts}
\usepackage{algorithmic}
\usepackage{graphicx}
\graphicspath{{figures/}}
\usepackage{textcomp}
\usepackage{xcolor}
\usepackage{booktabs}
\usepackage{multirow}
\usepackage[switch]{lineno}
\usepackage{hyperref}

\def\BibTeX{{\rm B\kern-.05em{\sc i\kern-.025em b}\kern-.08em
    T\kern-.1667em\lower.7ex\hbox{E}\kern-.125emX}}

\begin{document}

\title{ \vspace{-5mm} Transformer-guided Variable Projection for Phasor and Fault-Time Estimation Under DDC Offset\\
\vspace{-3mm}
}

\author{
\IEEEauthorblockN{ Marcus Chen I Wada, Sina Mohammadi, Van-Hai Bui, Wencong Su}
\IEEEauthorblockA{\textit{Department of Electrical and Computer Engineering} \\
\textit{University of Michigan-Dearborn}\\
Dearborn, USA\\
{wadamc, sinamo, vhbui, wencong}@umich.edu}
\vspace{-11mm}
}

\maketitle

\begin{abstract}
Decaying DC (DDC) offsets introduce substantial phasor estimation
errors during the first post-fault cycle. Recent approaches have sought interpretable waveform representations, but identifying analytical expressions can incur substantial computational costs. This paper proposes a
transformer-guided variable-projection estimator that extracts the
fundamental phasor, DDC parameters, and fault inception time
from point-on-wave (POW) samples as an interpretable analytical expression.
An offline-trained transformer provides the signal structure and
decay rate, while variable projection refines the parameters and
handles windows containing the fault without knowledge of time of fault.
Across single- and dual-DDC scenarios with time constants of
2.5--400~ms, the mean maximum overshoot (MO) is at most 2.7\% on post-fault windows and 2.2\% during online operation, where the DFT reaches 10--14\%.
The post-fault phasor settles within 5\% approximately 12.5~ms
after inception. Circuit-level validation on a 230-kV
two-source system yields 1.6\% mean MO,
compared with 9.5\% for the DFT.
\end{abstract}


\vspace{-4mm}
\section{Introduction}
Accurate and fast phasor estimation is fundamental to power-system protection, monitoring, and wide-area measurement applications. During faults, current and voltage waveforms may be distorted by harmonics, noise, frequency deviations, current-transformer effects, and exponentially decaying DC (DDC) components. Conventional Fourier-based estimators \cite{benmouyal2004synchronized} can exhibit considerable magnitude and phase errors following fault inception because DDC components violate their periodicity assumptions, potentially compromising protective-relay operation \cite{chang2022fault}.

\vspace{-1mm}
\subsection{Related Work}
\vspace{-1mm}
Recent research follows two complementary directions:
analyzing transient waveforms directly and improving
phasor estimation under DDC. Waveform measurement units
and point-on-wave (POW) recorders preserve synchronized
waveform samples \cite{MohsenianRad2023SynchroWaveforms},
enabling data-driven grid monitoring
\cite{Tong2024SynchroWaveformAI}. Beyond conventional
phasor measurements, these samples retain information
about fault inception, DDC dynamics, and harmonics,
which can be recovered through transient-aware modeling.

Analytical and signal-processing methods estimate the DDC within a fixed signal model: \cite{Liu2026DoubleDDC} solve the DFTs of sub-windows analytically for two DDCs, \cite{Rosolowski2026DDC} combines the DFT with a modified least-squares (LS) method based on notch-filter pairs, \cite{Chu2025EnhancedDFT} jointly estimate the frequency deviation and the DDC time constant from recursive DFT relations, and \cite{Ma2025MClass} and \cite{Wang2025DDC} reconstruct the DC offset iteratively by polynomial and cosine-series fitting, respectively. These methods are accurate when the waveform matches the assumed model, but each commits to a single model, and their accuracy degrades when the transient departs from it, for example with very short or multiple time constants \cite{Wang2025DDC} or strong transients right after fault inception \cite{Rosolowski2026DDC}.

Machine learning methods offer an alternative. The network of \cite{daSilva2015NN} is in effect an adaptive LS fit whose weights, trained online on each window, are the coefficients of harmonic sinusoids and a Taylor-approximation DDC term. 
Offline trained networks map sub-cycle windows either to the undistorted waveform, from which the phasor is fitted by LS \cite{Sok2022DNN}, or directly to the fundamental amplitude and phase under two DDCs and frequency deviations \cite{Mohammadi2024NN}, while \cite{Tajdinian2025SVM} reconstructs current-transformer waveforms with support vector machine and LS support vector machine models before a DDC robust modified DFT. These estimators fit a fixed, approximate model or learn a black-box mapping, and none identifies the analytical structure of the fault current.

    Deep symbolic regression (DSR) enables interpretable
fault-current estimation \cite{mohammadi2026symbolicphasor} and has been used to
identify the harmonic content of grid voltages for converter control \cite{hollweg2023rmrac}, but requires computationally expensive symbolic searches
on each window. Conversely, LS methods are
efficient but require predefined signal structures and
nonlinear parameters. Inspired by SymFormer's joint
structure-and-constant prediction \cite{vastl2024symformer}, this work uses a transformer to initialize a variable projection fit \cite{golub2003separable}, avoiding online symbolic search while retaining an interpretable waveform representation. The main contributions are:
\vspace{-1mm}

\begin{itemize}
\item Single-pass identification of fault-current components and initialization of coefficients for subsequent physics-constrained refinement.
\item A variable-projection refinement with a continuous sigmoid switch between pre- and post-fault models, which estimates the fault time jointly with the decay rates, without requiring prior knowledge of the fault occurrence time.
\item An offline-trained estimator requiring no online training
or symbolic search, with per-window execution time below
the quarter-cycle update interval on a single CPU thread,
validated on multiple scenarios and
electromagnetic-transient simulations.
\end{itemize}

\vspace{-2mm}
\section{Proposed Method}

\subsection{Problem Formulation}

Under fault conditions, current waveforms may contain harmonics
and one or more DDC components
\cite{haji2024novel}, modeled as
\vspace{-3mm}

\begin{equation}
i(t) =
\begin{cases}
I_{pre}\sin(\omega t+\phi_{pre}),
& t<t_f, \\[6pt]

\begin{aligned}
I_{1}\sin(\omega t+\phi_{1})
&+\displaystyle\sum_{k\in\mathcal{H}}
I_k\sin(k\omega t+\phi_k) \\[-2pt]
&+\displaystyle\sum_{j=1}^{N_d}
D_j e^{-(t-t_f)/\tau_j},
\end{aligned}
& t\geq t_f.
\end{cases}
\label{eq:fault_waveform}
\end{equation}

where $t_f$ is the fault inception time, $\omega=2\pi f$, $\omega_0$ is the
nominal angular frequency,
$\mathcal{H}$ is the set of harmonic orders, and $N_d$ is the
number of DDC components. The parameters $I_k$ and $\phi_k$
denote sinusoidal amplitudes and phases, while $D_j$ and
$\tau_j$ represent DDC amplitudes and time constants.

The sampled waveform is
$y[n]=i(nT_s)+\eta[n]$, where $T_s=1/f_s$ and the noise,
$\eta[n]\sim\mathcal{N}(0,\sigma^2)$. Given the measured sequence $\{y[n]\}_{n=0}^{N-1}$,
the objective is to recover the fundamental phasor
$(I_1,\phi_1)$ and, when present, the fault inception
time $t_f$. To represent the possible waveform components,
the admissible signal structure is defined as
\begin{equation}
\mathcal{S}_{\mathrm{adm}}
=\{F,\mathrm{DDC}_1,\mathrm{DDC}_2,H_3,H_5\},
\label{eq:model_assembly}
\end{equation}
where $F$ denotes the fundamental and $H_k$ the $k$-th
harmonic. The estimator selects the relevant components
from this set for subsequent refinement.

\vspace{-2mm}
\subsection{One-Shot Structure and Parameter-Predicting Transformer}

\textit{Input representation:} Each window is peak-normalized as $x[n]=y[n]/\max_k|y[k]|$, since the structure and decay rate used by the subsequent refinement are independent of scale. 
Each sample is represented by $[x[n],\sin(\omega_0 nT_s),\cos(\omega_0 nT_s)]$
and linearly embedded into $\mathbb{R}^{64}$. The nominal-frequency encoding, with $\omega_0$ the nominal angular frequency, provides sample position and phase information, allowing the encoder to distinguish the fundamental from transient components. A learned classification token is offset by a linear projection of  $[10^3 t_0, \sin(\omega_0t_0), \cos(\omega_0t_0)]$, where $t_0$ is the window start time (s) measured from the detected fault inception. These features provide a phase reference and window position within the DDC.

\textit{Architecture and outputs:}
Three transformer encoder layers ($d=64$, four attention heads,
128-unit feed-forward dimension) outputs a classification-token
representation shared by two heads. The structure head selects waveform components $\mathcal{S}\subseteq\mathcal{S}_{\mathrm{adm}}$ using a probability threshold, with the fundamental always included.
A two-layer coefficient head predicts
$[\hat a_1,\hat b_1,\hat D,\hat{\tilde{\lambda}},
\hat a_3,\hat b_3,\hat a_5,\hat b_5]$,
where $a_k=I_k\cos\phi_k$ and $b_k=I_k\sin\phi_k$. The normalized prediction $\hat{\tilde{\lambda}}$
is mapped to the physical decay rate $\hat{\lambda}$.
Only the selected structure $\mathcal{S}$ and
$\hat{\lambda}$ are passed to variable projection,
which refines the decay rate, recomputes the
coefficients, and returns the final phasor.

\begin{figure}[!t]
    \centering
    \includegraphics[width=0.95\columnwidth]{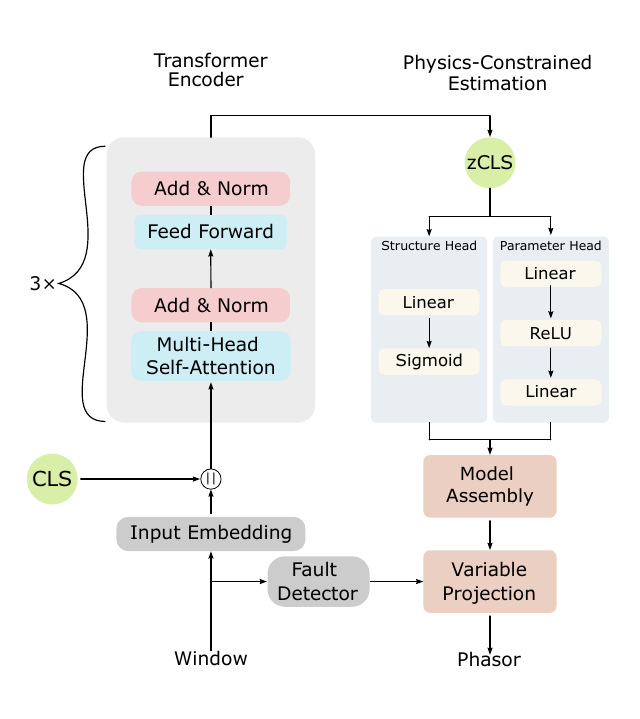}
    \vspace{-10mm}
    \caption{Proposed estimator architecture.}
    \label{fig:architecture}
    \vspace{-5mm}
\end{figure}

\subsection{Physics-Constrained Refinement}
\label{sec:refine}

For a given structure $\mathcal{S}$ and decay rates
$\boldsymbol{\lambda}$, the waveform model is linear in its remaining
coefficients $\boldsymbol{\theta}:=
[\{a_k,b_k\}_{k\in\{1\}\cup\mathcal{H}},
\{D_j\}_{j=1}^{N_d}]^T$. The design matrix $\boldsymbol{\Phi}$ contains
sine--cosine column pairs at $k\omega_0$ for the fundamental and the
selected harmonics, and a column $e^{-\lambda_j\mathbf{t}}$ for each DDC component.
The coefficients are obtained in closed form as
$\hat{\boldsymbol{\theta}}=\boldsymbol{\Phi}^{+}\mathbf{y}$, while the
nonlinear parameters are refined by minimizing the variable-projection
residual \cite{golub2003separable}
\begin{equation}
J=\|\mathbf{r}\|_2^2,\qquad
\mathbf{r}=\bigl(\mathbf{I}-\boldsymbol{\Phi}\boldsymbol{\Phi}^{+}\bigr)\mathbf{y}.
\label{eq:varpro}
\end{equation}
For windows containing the fault, a continuous transition between the
pre- and post-fault models is introduced:
\begin{equation}
\begin{aligned}
y[n] &\approx (1-s[n])y_{\mathrm{pre}}[n]
+s[n]y_{\mathrm{post}}[n],\\
s[n] &= \frac{1}{1+e^{-(n-m)/w}}.
\end{aligned}
\label{eq:split_model}
\end{equation}
where $m\in\mathbb{R}$ is the fault position relative to
the window start $t_w$, in samples
($t_f=t_w+mT_s$), $y_{\mathrm{pre}}$ is a fundamental sinusoid, and
$y_{\mathrm{post}}$ is the post-fault model with structure $\mathcal{S}$.
The width $w=0.375$ samples gives a transition of about 1.5 samples,
close to a step, while keeping the residual differentiable in $m$
between samples; for the same reason, the DDC terms $e^{-\lambda_j c[n]}$
use the smooth clock $c[n]=T_s\ln(1+e^{n-m})$
instead of $T_s\max(n-m,0)$. For fixed
$(m,\boldsymbol{\lambda})$, \eqref{eq:split_model} remains linear in the
pre- and post-fault coefficients, so the fault time and the decay rates
are refined jointly by the same variable projection.

The nonlinear parameters are the log-decay rates
$\mathbf{u}=\ln\boldsymbol{\lambda}$, augmented by
the fault position $m$ in fault-containing windows:
$\boldsymbol{\alpha}=[m,\mathbf{u}^T]^T$.
They are refined using Gauss--Newton with Kaufman's
variable-projection Jacobian approximation:

\begin{equation}
\begin{gathered}
(\mathbf{K}^T\mathbf{K})\Delta\boldsymbol{\alpha}
= -\mathbf{K}^T\mathbf{r},\\
\mathbf{K}_i \approx
-(\mathbf{I}-\boldsymbol{\Phi}\boldsymbol{\Phi}^{+})
\frac{\partial\boldsymbol{\Phi}}{\partial\alpha_i}
\hat{\boldsymbol{\theta}}.
\end{gathered}
\label{eq:kaufman_jacobian}
\end{equation}

Steps are bounded by $|\Delta u_j|\le\ln2$ and $|\Delta m|\le6$ samples,
with $\lambda_j\in[2.5,400]~\mathrm{s}^{-1}$, and accepted only if they
reduce $J$. Post-fault windows start from the transformer's
$\hat\lambda$, with five iterations, selected through convergence experiments. Windows containing the fault use Levenberg--Marquardt \cite{golub2003separable} damping, with
eight iterations at the first detection and three in later windows,
which start from the previous estimates. The refined $m$ is then rounded
to the neighboring sample with the lowest hard-switch residual, which
gives $\hat t_f$, and the window is refitted with the hard switch and the
clock $\max(n-m,0)\,T_s$. The optimized decay rates $\boldsymbol{\lambda}^{*}$
are used to recompute the linear coefficients
$\hat{\boldsymbol{\theta}}=
\boldsymbol{\Phi}(\boldsymbol{\lambda}^{*})^{+}\mathbf{y}$.
The fundamental phasor is then recovered as
$\hat I_1=\sqrt{\hat a_1^2+\hat b_1^2}$ and
$\hat\phi_1=\operatorname{atan2}(\hat b_1,\hat a_1)$.

\vspace{-3mm}
\subsection{Online Estimation}
\label{sec:online}

A one-cycle window slides over the measurements with a quarter cycle step. The transformer is trained offline, while only the estimated fault time and decay rates are carried between windows. The model uses its full structure, and fault containing windows are fitted
using~\eqref{eq:split_model} without prior $t_f$ knowledge. The overall methodology is illustrated in Fig.~\ref{fig:architecture}.

\textit{Fault detection:}
For a periodic pre-fault signal, the cycle difference
$d[n]=y[n]-y[n-N]$ ideally contains only noise.
A fault is declared at the first of three consecutive samples
satisfying $|d[n]|>6\hat\sigma_d$, where
$\hat\sigma_d=1.4826\,\mathrm{MAD}(d)$; the median absolute deviation of $d$ over the preceding cycle. This robust threshold adapts to measurement
noise while limiting sensitivity to isolated transients.
Before detection, the estimator reports the LS
fundamental phasor. The detected sample initializes $m$.

\begin{figure}[!t]
    \centering
    \includegraphics[width=0.8\columnwidth]
    {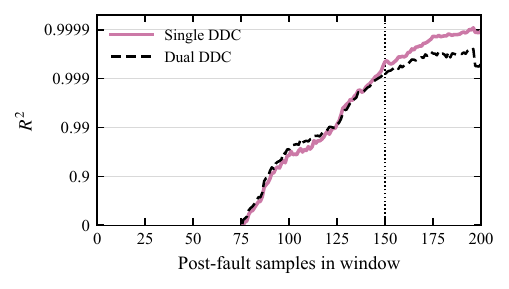}
    \vspace{-5mm}
    \caption{Reconstruction $R^2$ versus the number of
    post-fault samples for single- and dual-DDC signals.}
    \label{fig:postsamples}
    \vspace{-5mm}
\end{figure}

\textit{Reporting:}
The pre-fault phasor is retained until 150 post-fault samples
are available. Thereafter, the post-fault phasor is estimated
with $m$ from detector and the transformer initialized on the
post-fault segment ($t_0=0$).
For fully post-fault windows, $t_0=t_w-\hat t_f$,
and subsequent windows reuse the previously refined
decay rates. 
As shown in Fig.~\ref{fig:postsamples}, 150 samples
were selected as a practical reporting threshold,
at which $R^2$ exceeds 0.999 for both DDC cases.

\vspace{-3mm}
\section{Training and Evaluation}

\subsection{Signal Generation Model}

Synthetic waveforms are generated using~\eqref{eq:fault_waveform}
at 12~kHz over 200-sample windows, with fundamental amplitudes
of 1--20~p.u., random phases, and 20--40~dB SNR.
The DDC time constants span 2.5--400~ms. 

Training windows begin uniformly within the first 40~ms
after fault inception and contain only post-fault samples.
Windows containing pre-fault data are handled by the
online procedure in Section~\ref{sec:online}.

\subsection{Training Procedure}

The network is trained with a multi-task loss combining
binary cross-entropy (BCE) for structure classification
and mean squared error (MSE) for coefficient regression,

\begin{equation}
\mathcal{L}=\frac{1}{4}\sum_{k}\mathrm{BCE}(p_k,s_k)
+\frac{\sum_{i} w_i(\hat r_i-r_i)^2}{\sum_i w_i},
\end{equation}
where \(s_k\in\{0,1\}\) indicates the presence of term \(k\), \(r_i\) are
the normalized coefficient targets, and the mask \(w_i\) removes the
coefficients of absent terms from the regression loss. The model is trained using 4\,000 training and 800 validation windows. Decay rates are sampled log-uniformly. Training uses Adam with a learning
rate of $10^{-3}$ for up to 20 epochs, keeping the checkpoint with the lowest validation loss.

\subsection{Baseline Estimators}

The proposed method is compared with three baselines evaluated on the
same records: (i) the full-cycle DFT, on which relay phasor estimation is
based \cite{benmouyal2004synchronized}; (ii) the closed-form DDC removal method by Abdoos \cite{abdoos2016accurate}, which estimates a
single DDC from the cycle integral and the endpoint difference before a
DFT of the residual; and (iii) PySR, our replication of the deep symbolic
regression search of \cite{mohammadi2026symbolicphasor}, run on each
window.

\subsection{Evaluation Metrics}

Performance is evaluated using the coefficient of determination
$R^2$ between the true and reconstructed fundamental waveforms,
MO and total vector
error (TVE):
\begin{align}
\mathrm{MO} &=
100\,\frac{\max_t \hat{I}_1(t)-I_1}{I_1}, \\
\mathrm{TVE} &=
100\,\frac{|\hat{\mathcal{I}}_1-\mathcal{I}_1|}
{|\mathcal{I}_1|}, \quad \mathcal{I}:= \frac{I_1}{\sqrt{2}}e^{j\phi_1} .
\end{align}

Three scenario families are considered: low-$\tau$ single DDC ($\tau$ = 2.5--10~ms), extended-$\tau$ single DDC ($\tau$ = 50--200~ms), and dual DDC ($\tau_2$ up to 400~ms).

For post-fault evaluation, 50-ms records are processed using
one-cycle windows with a quarter-cycle step over 30 noise
realizations. Overlapping reconstructions are averaged;
$R^2$ and MO are computed on the averaged trajectory,
while TVE is evaluated per window.

For online evaluation, windows advance by quarter cycle without
averaging or known fault time. MO includes all post-inception estimates,
while settling time is defined as the time after which the
estimated amplitude remains within 5\% or 2\% of its true value.

\vspace{-3mm}
\section{Results}

\begin{figure*}[!t]
    \centering
    \includegraphics[width=0.85\textwidth]{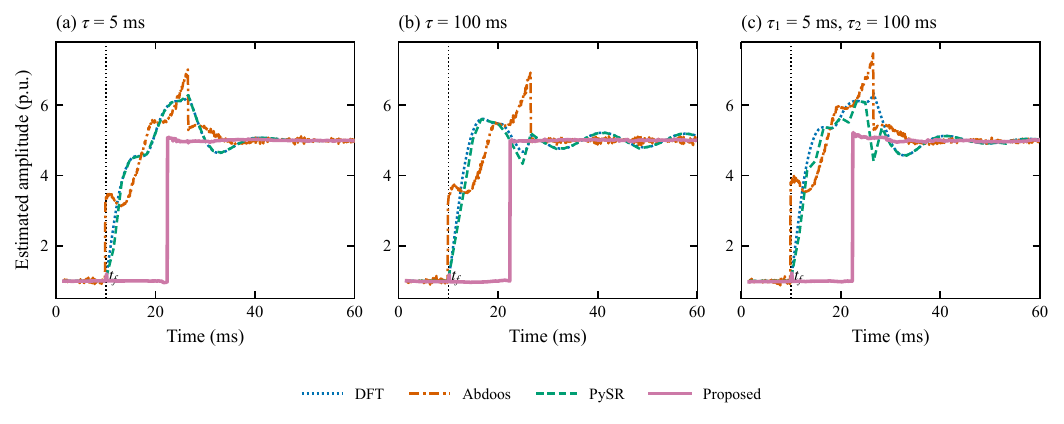}
    \vspace{-5mm}
    \caption{Online fundamental amplitude estimation with a
    quarter cycle step and without prior knowledge of fault inception
    time. The proposed method retains the pre-fault phasor until
    150 post-fault samples are available.}
    \label{fig:amplitude_time}
    \vspace{-5mm}
\end{figure*}

\begin{table}[!t]
\centering
\caption{Post-Fault Phasor Estimation Accuracy}
\label{tab:dco_scenario_comparison}
\renewcommand{\arraystretch}{1.08}
\setlength{\tabcolsep}{5pt}
\scriptsize
\begin{tabular}{llccc}
\toprule
\textbf{Scenario} & \textbf{Method}
& $\boldsymbol{R^2}$
& \textbf{MO (\%)}
& \textbf{TVE (\%)} \\
\midrule
\multirow{4}{*}{Low-$\tau$}
& DFT & 0.7495 & 59.82 & 27.48 \\
& Abdoos & 0.9481 & 32.24 & 14.59 \\
& PySR & 0.7049 & 55.44 & 29.34 \\
& Proposed & \textbf{0.9998} & \textbf{1.56}
& \textbf{1.46} \\
\midrule
\multirow{4}{*}{Ext.-$\tau$}
& DFT & 0.9794 & 13.15 & 22.00 \\
& Abdoos & 0.9964 & 3.84 & 5.96 \\
& PySR & 0.9758 & 5.68 & 18.80 \\
& Proposed & \textbf{0.9986} & \textbf{2.69}
& \textbf{3.56} \\
\midrule
\multirow{4}{*}{Dual DDC}
& DFT & 0.8900 & 39.19 & 29.04 \\
& Abdoos & 0.9954 & 6.76 & 6.91 \\
& PySR & 0.8831 & 32.42 & 27.32 \\
& Proposed & \textbf{0.9996} & \textbf{1.82}
& \textbf{2.26} \\
\bottomrule
\end{tabular}

\vspace{-3mm}
\end{table}

\subsection{Synthetic Data}

The proposed method is evaluated on synthetic waveforms under two settings, both with SNR uniformly sampled from 20 to 40~dB. First, 30 fully post-fault windows per scenario, with independent noise realizations, are used to assess phasor estimation accuracy. Second, continuous online records are used to evaluate transient behavior, fault-time estimation, and settling performance without prior knowledge of fault inception time.

\subsubsection{Post-Fault Windows}
Table~\ref{tab:dco_scenario_comparison} shows that baseline performance depends strongly on DDC dynamics. The largest differences occur at short time constants, where the exponential decays substantially within a single estimation window. In this regime, Abdoos' first-order approximation becomes less representative of the DDC, while the PySR search is dominated by the fundamental component, performing similarly to DFT.  In contrast, the proposed method explicitly retains the exponential term and refines its decay rate
through variable projection, substantially reducing transient errors.

As the DDC time constant increases, the exponential varies more slowly within each window, improving baseline performance and narrowing the gap with the proposed method, which nevertheless remains the most accurate. Under dual-DDC conditions, the proposed method maintains accuracy comparable to that achieved in the extended-time-constant single-DDC scenario.

\begin{table}[!t]
\centering
\caption{Online Phasor Estimation and Settling Time}
\label{tab:online}
\renewcommand{\arraystretch}{1.08}
\setlength{\tabcolsep}{3.5pt}
\scriptsize
\begin{tabular}{llccc}
\toprule
\textbf{Scenario} & \textbf{Method}
& \shortstack{\textbf{MO (\%)}\\Mean / p90}
& \shortstack{\textbf{TVE (\%)}\\Mean}
& \shortstack{\textbf{Settling (ms)}\\5\% / 2\%} \\
\midrule
\multirow{3}{*}{Single DDC}
& DFT & 10.20 / 21.41 & 6.24 & 19.7 / 29.2 \\
& Abdoos & 13.76 / 36.89 & 2.69 & 18.3 / 29.6 \\
& PySR & 10.53 / 21.86 & 6.76 & 24.3 / 29.5 \\
& Proposed & \textbf{1.32 / 2.25} & \textbf{0.60} & \textbf{12.4 / 13.6} \\
\midrule
\multirow{3}{*}{Dual DDC}
& DFT & 13.59 / 26.29 & 9.20 & 24.6 / 31.1 \\
& Abdoos & 18.03 / 45.83 & 4.38 & 20.6 / 29.4 \\
& PySR & 13.04 / 26.03 & 9.39 & 24.8 / 31.1 \\
& Proposed & \textbf{2.22 / 3.65}
& \textbf{0.88} & \textbf{12.6 / 15.7} \\
\midrule
\multirow{3}{*}{Circuit}
& DFT & 9.50 / 15.84 & 4.30 & 24.5 / 33.4 \\
& Abdoos & 3.30 / 5.57 & 1.72 & 21.4 / 35.1 \\
& PySR & 9.97 / 16.45 & 5.08 & 30.2 / 36.6 \\
& Proposed & \textbf{1.62 / 3.12}
& \textbf{0.60} & \textbf{13.0 / 15.8} \\
\bottomrule
\end{tabular}

\vspace{-5mm}
\end{table}

\subsubsection{Online Operation and Fault-Time Estimation}

Fig.~\ref{fig:amplitude_time} illustrates the online behavior of the estimators. While baseline methods exhibit large transient excursions as the window crosses the fault, the proposed method maintains the pre-fault phasor until sufficient post-fault information is available, as detailed in \ref{sec:online}. By estimating fault inception within variable projection, the method avoids fitting a post-fault model to mixed pre- and post-fault samples, substantially reducing transient errors and achieving faster settling. This advantage persists under dual-DDC conditions as shown in Table~\ref{tab:online}.

Fault-time estimation, described in Section~\ref{sec:online},
was evaluated on 250 single-DDC records ($\tau=50$~ms, 30~dB SNR), with post-fault fundamental amplitudes $I_1$ ranging from 1.1 to 5~pu and initial DC fractions $D/I_1$ ranging from 0 to 100\%, using 10 records per case. Faults with $I_1\geq1.5$~pu or $D/I_1\geq50\%$ were reliably detected, with a median timing error of one sample. Missed detections occurred only for weaker faults ($I_1\leq1.2$~pu and $D/I_1\leq20\%$), where limited waveform changes make fault identification more difficult. No false alarms were observed, and MO remained below 4.1\% across all tested records.

The proposed method requires 2.4–2.5 ms per window on a single CPU thread, below the 4.17-ms quarter-cycle update interval. 

\vspace{-3mm}
\subsection{Circuit-Level Validation}

To evaluate the estimator beyond the synthetic signal model, adapted from \cite{haji2024novel}, single phase faults were simulated on a three-phase,
230-kV, 60-Hz two-source system in MATLAB/Simscape Electrical. Fault resistance, location, and inception angle were varied across 270 records. The resulting currents include circuit-dependent DC offsets and additional dynamics introduced by line capacitance, zero-sequence paths, and anti-aliasing filtering, which are not explicitly represented in the assumed waveform model. All records were processed online without retraining.

As shown in table~\ref{tab:online} , the proposed method maintains low 
MO and TVE, with substantially faster settling than baseline models, demonstrating generalization to circuit-level simulations despite dynamics not explicitly captured by the assumed signal model. Fault inception was successfully detected in 265 of 270 records, with missed detections occurring only at low SNR (below 21 dB).

\subsection{Sensitivity to Frequency Deviation}

Frequency sensitivity was evaluated over 59.5--60.5~Hz, with all estimators fixed at the nominal 60~Hz. As shown in Table~\ref{tab:offnominal}, the proposed method consistently achieves the lowest mean TVE under both single- and dual-DDC conditions. Its error increases approximately symmetrically with frequency deviation, indicating sensitivity to mismatch in the fixed-frequency sinusoidal basis. The baseline methods exhibit smaller changes but maintain substantially higher absolute errors. The greater sensitivity under dual DDC suggests that additional exponential components complicate phasor estimation under model mismatch. These results motivate incorporating frequency as an additional nonlinear parameter in variable projection, which are out of the scope of current work.

\begin{table}[!t]
\centering
\caption{Mean Online TVE (\%) Under Frequency Deviation}
\label{tab:offnominal}
\scriptsize
\setlength{\tabcolsep}{4pt}
\begin{tabular}{c cccc cccc}
\toprule
& \multicolumn{4}{c}{Single DDC}
& \multicolumn{4}{c}{Dual DDC}\\
\cmidrule(lr){2-5}\cmidrule(lr){6-9}
$f$ (Hz) & DFT & Abdoos & PySR & Prop.
& DFT & Abdoos & PySR & Prop.\\
\midrule
59.5 & 6.25 & 2.93 & 6.84 & \textbf{1.44}
     & 9.21 & 4.53 & 9.46 & \textbf{1.92}\\
60.5 & 6.25 & 2.94 & 6.79 & \textbf{1.47}
     & 9.21 & 4.55 & 9.50 & \textbf{1.96}\\
\bottomrule
\end{tabular}
\vspace{-5mm}
\end{table}

\vspace{-2mm}
\section{Conclusion}

A transformer trained offline can supply the structure and the
nonlinear starting point of a physics-constrained variable-projection
fit, turning one window of POW samples into an interpretable expression
of the fault current: the fundamental phasor, the DDC parameters, and,
for windows that contain the fault, the fault time. Without known fault instant, the estimator kept the mean maximum overshoot through the fault at
1.3--2.2\%, where the DFT reached 10--14\%, reported the post-fault phasor
about 12.5~ms after inception, and transferred without retraining to
fault currents from an electromagnetic-transient simulation. The limitations are the
nominal-frequency model, a structure head
that is only preliminarily tested, and simulated data. Future work will
add frequency to the variable projection, train on variable structures,
and evaluate field POW records.

\vspace{-2mm}
\bibliographystyle{IEEEtran}
\bibliography{Ref}

\end{document}